\documentclass[12pt]{article}
\usepackage{graphicx}
\usepackage{cite}
\usepackage{amsmath}

\begin{document}
\title{To understanding of the mechanisms of DNA deactivation in ion therapy of cancer cells}%
\author{D.V. Piatnytskyi, O.O. Zdorevskyi, S.M. Perepelytsya, S.N.
Volkov\\
Bogolyubov Institute for Theoretical Physics, NAS of
Ukraine,\\14-b Metrolohichna Str., Kyiv, 03680, Ukraine \\
snvolkov@bitp.kiev.ua }\maketitle

\setcounter{page}{1}%
\maketitle

\begin{abstract}
Changes in the medium of biological cell nucleus under
ion beam action is considered as a possible cause of cell
functioning disruption in the living body.
As the most long-lived molecular product appeared
in the cell after the passage of high energy ions, the
hydrogen peroxide molecule is picked out. The possibility of the
formation of stable complexes of hydrogen peroxide molecules 
with active sites of DNA nonspecific recognition (phosphate groups
of the double helix backbone) is studied, and the formation of
stable DNA-peroxide complexes is considered. Due to the negative
charge on the oxygen atoms of DNA phosphate group in solution
the counterions that under natural conditions neutralize
the double helix have been also taken into consideration.
The complexes consisting of oxygen atoms of DNA
phosphate group, H$_2$O$_2$ and H$_2$O molecules, and Na$^{+}$
counterion have been considered. Energy of the complexes have been
determined based on the electrostatic and van der Waals interactions
within the approach of atom-atom potential functions.
The stability of various configurations of molecular
complexes has been estimated. It has been found that hydrogen
peroxide molecules can form the stable complexes with phosphate
groups of DNA and counterions which are no less stable than the
complexes with water molecules. It is shown that the formation of stable
complexes of H$_2$O$_2$--Na$^{+}$--PO$_{4}^{-}$ can be detected
experimentally by the observation of specific DNA vibrations in the
low-frequency Raman spectra. The interaction of H$_2 $O$_2$
molecule with phosphate group of the double helix backbone can
block the processes of DNA biological functioning and induce the
deactivation of the genetic apparatus of the cell. Thus, the
new channel of high-energy ions action on living cell has
been proposed.
\end{abstract}

\section{Introduction}
\label{intro}

Invention of new instruments for cancer therapy is known to be one
of the most important challenges of modern science. In spite of
the increase of funding for the development of new anticancer
drugs the illness rate is still very high~\cite{NatOutl}. In this
regard, the elaboration of non pharmacological methods of cancer
therapy are stimulated. The last decades, use of high energy ion
beams has established itself as an effective treatment of cancer
disease. This method is based on well-known Bragg
effect~\cite{Bregg,Kraft,Suit,Schlaff}. Heavy ions (usually
protons or $^{12}$C$^{6+}$), accelerated to the energies of
hundreds of MeV, are targeted into tumor. Within organism tissue
the ions lose their initial energy mainly at the end of their
trajectory of motion, and the energetic peak is formed (the Bragg
peak). The initial beam energy is chosen in such a way that the
position of the Bragg peak coincide with the location of tumor in
human body (it can be 5-10 cm from body surface). The energy which
is lost in living tissues destroys cancer cells and consequently
the whole tumor. The method of ion beam therapy is much more
effective than X-ray therapy, because of local action in the Bragg
peak, especially in the case of places of difficult access, such
as human brain~\cite{Kraft,Suit,Schlaff}. In spite of great
practical interest and large number of ion therapy facilities
already built, there is no clear understanding of the molecular
mechanism of action of high energy ions.

Starting with the early study~\cite{TR} the action of radiation on
living organisms is known to be related with the DNA damage. The
breaking of DNA double helix affects the mechanism of storage and
transfer of genetic information which leads to cell death. Under
heavy ion irradiation the breaks of hydrogen bounds in nucleotide
pairs occur, which cause DNA melting, single and double strand
breaks in double helix~\cite{Kraft,Suit,Schlaff}. Single strand
breaks in DNA and melting can be repaired due to the complementary
structure of double helix and cell reparation mechanisms. The
double strand breaks in DNA lead to damage of macromolecule, which
can not be repaired ~\cite{Saenger}. Thus the formation of large
number of double strand breaks of double helix is necessary for
deactivation and destruction of DNA molecule.

The double strand breaks in DNA macromolecule occur due to the
large quantity of secondary electrons born in the medium under the
action of ion beam~\cite{Boudaoiffa,Hamada,Sol09,Sol13}. Some
molecules decay into atoms and radicals that can break chemical
bonds in biological macromolecules. At the same time, the
calculations show that the number of the double strand breaks,
made by secondary electrons and radicals, is insufficient for cell
death ~\cite{Sol09}. Therefore, additional channels of ion beam
action on living tissues should exist.

To describe DNA damage the formation of the shockwave in the
medium after passage of high energy ions is
suggested~\cite{Shok1,Shok2}. Within the framework of shockwave
mechanism the destruction of DNA double helix  may occur even in
the case of macromolecules situated far from ion hit point in the
cell. The molecular dynamics simulations shows that shockwave
cause large number of double strand breaks in DNA double helix
~\cite{Sol13}. However, in spite of high efficiency of proposed
mechanism, there are processes which can prevent DNA damage in the
living cell, particularly, the histone proteins restraining the
macromolecule in frames of chromatin.

The deactivation of DNA macromolecule may be induced by changes of
medium that occur after the passage of high energy ions. The goal
of the present work is to consider the changes of medium in the
Bragg peak region and to pick out new molecular compounds that can
influence the structure and dynamics of DNA double helix. The
analysis of changes in water medium  induced by ion beams is
carried out in second section and the most probable molecular
compounds formed in the Bragg peak are analyzed. The third section
is dedicated to the description of calculation method of
interaction of appeared molecules and atoms of DNA double helix.
The results of energy calculations are presented in fourth section
and the most stable complexes are selected. In the fifth section
the low-frequency vibrations of formed complexes are studied, and
the specific modes of the complexes are found in the Raman spectra
of DNA. The possible action of medium changes on DNA biological
functioning is discussed in the sixth section.

\section{Medium change under ion beam radiation}
\label{sec:1} The ions of high energy passing through intracellular medium
induce the process of decay of water molecules (water radiolysis),
different chemical and nuclear
reactions~\cite{Pshenichnov,Haettner,Soltani,Pastina,Caer,Wasselin,Kreipl,Uehara}.
The processes of nuclear fragmentation are showed be not
significant for environment~\cite{Haettner,Soltani}, whereas the
fragmentation of water molecules change essentially the
composition and properties of water
medium~\cite{Kraft,Suit,Schlaff,Sol09}. In the process of water
radiolysis the secondary electrons ($e^{-}$), radicals
(OH$^{\bullet}$ and HO$_2^{\bullet}$), molecules (H$_2$O$_2$ and
H$_2$O), and other products (H$^{\bullet}$, OH$^{-}$,
H$_3$O$^{+}$) are produced ~\cite{Pastina,Caer,Wasselin}.

\begin{figure}
\begin{center}
\resizebox{0.5\textwidth}{!}{%
  \includegraphics{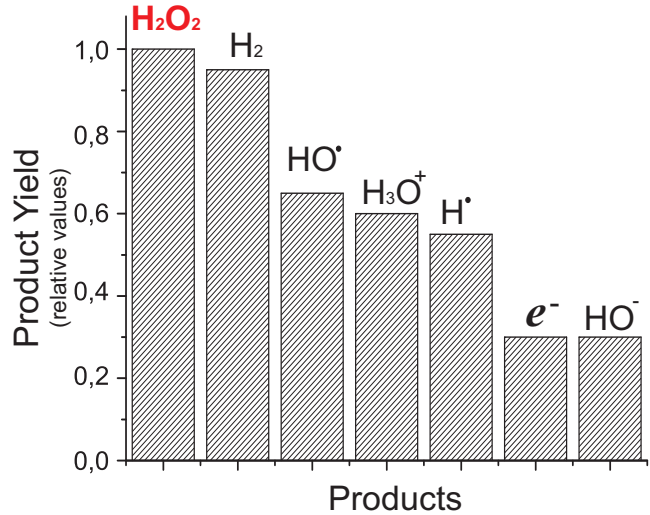}
}
\caption{Products of water radiolysis after $10^{-6}$ s of ion
passage. Bar graph is built using the Monte Carlo simulation
data~\cite{Kreipl}.}
\label{fig:1}       
\end{center}
\end{figure}

Quantity and composition of radiolysis products depend on absorbed
radiation dose.  At small doses the number of secondary electrons
and radicals is the largest,  while the increase of the dose
energy leads to the increase of molecular products yields. Due to
the high energy transfer  the number of molecular products is
expected to be the largest in the Bragg peak region. The Monte
Carlo simulation data for water radiolysis~\cite{Kreipl,Uehara}
show that in the Bragg peak the secondary electrons and hydroxyl
radicals yields prevail at early stage of radiolysis (time
10$^{-12}$ s). After the time 10$^{-6}$ s the reactions
OH$^{\bullet}+$OH$^{\bullet}\rightarrow$H$_2$O$_2$,
HO$_2^{\bullet}+$HO$_2^{\bullet}\rightarrow$H$_2$O$_2+$O$_2$
occur, and the yields of molecular products  become larger than
other yields~\cite{Kreipl}. Taking into consideration that the
microsecond time scale is typical for biological processes, the
equilibrium composition of the intracellular medium after the
passage of high energy ions must be enriched by the molecular
products of water radiolysis.

To analyze the composition of the radiolysis products at the Bragg
peak the Monte Carlo simulation data~\cite{Kreipl} are used. The
distribution of radiolysis products at the time 1 $\mu$s after ion
passage are showed in the Figure 1. It is seen that the number of
hydrogen peroxide molecules in intracellular medium is larger than
other products of water radiolysis. The number of H$_{2}$
molecules is also considerable, but these molecules don't have
high reactivity and as a result do not threaten biological
macromolecules. Radicals which also have substantial yields after
some time transform into neutral water molecules or hydrogen
peroxide. Thus, the composition of intracellular medium changes
after ion beam irradiation  and significant amounts of hydrogen
peroxide molecules appear.

The concentration of hydrogen peroxide in the living cell is kept
constant by special ferments ~\cite{Watson}, but it may be
increased greatly by the action of ion beams. Increasing the
amount of hydrogen peroxide molecules raises the probability of
their interaction with DNA. The interaction of large number of
H$_2$O$_2$ molecules and DNA can deactivate the cell genetical
apparatus and induce cell death. Therapeutic effect of the
hydrogen peroxide has been found under the treatment of cancer
disease (see for example~\cite{Manda}). Ion therapy allows to
increase the concentration of hydrogen peroxide in tumor cells
with high precision and with minimal damage of healthy tissues.

The hydrogen peroxide molecule can interact with DNA molecule via
charged atomic groups of the double helix. It is well known that
the phosphate groups of double helix backbone bear the highest
charge~\cite{Saenger}. Under the natural conditions it is
neutralized by metal ions (for example, Na$^{+}$ or K$^{+}$).
Thus, the interaction of H$_2$O$_2$ molecule with the phosphate
groups of the double helix and tethered metal counterions is
expected to be the most probable.

\section{Model and method of energy calculation}
\label{sec:2}

The capability of hydrogen peroxide to interact with DNA is
studied by calculating the energy of different complexes. The
considered complexes consist of DNA phosphate group, hydrogen
peroxide molecule, water molecule and Na$^{+}$ counterion. For the
determination of relative stability of different complexes two
oxygen atoms of the phosphate group are sufficient to consider.
Other atoms of PO$_{4}^{-}$ are essential for accurate
calculations of energy. Schematic constructions and structural
parameters of molecules in complexes are showed at Figure 2.

\begin{table*}
\begin{center}
\noindent\caption{Parameters of potentials.
}\vskip3mm\tabcolsep4.5pt
\noindent{\footnotesize\begin{tabular}{lcc|lc|lcc}
\multicolumn{3}{c}{Atomic charges in
molecules}&\multicolumn{2}{c}{Potential (4)
~\cite{PV2}}&\multicolumn{3}{c}{Potential (3)
~\cite{Poltev1,Poltev2}}\\\hline
Molecule&H&O&Parameter&Na$^{+}$&Parameter&O--O&O--H\\\hline
H$_{2}$O&+0.33 ($e$)&-0.66 ($e$)&q ($e$)&+1.00&$A_{ij}$ (\AA$^{6}$kcal/mol)&200&86\\
H$_{2}$O$_{2}$&+0.41 ($e$)&-0.41 ($e$)&$r_{0}$ (\AA)&2.35&$B_{ij}$ (\AA$^{12}$kcal/mol)&12900&31300\\
PO$_{4}^{-}$&-&-0.50 ($e$)&$b$ (\AA)&0.3&&&\\\hline
\end{tabular}}
\end{center}
\end{table*}

\begin{figure}[b!]
\centering
\resizebox{0.7\textwidth}{!}{%
  \includegraphics{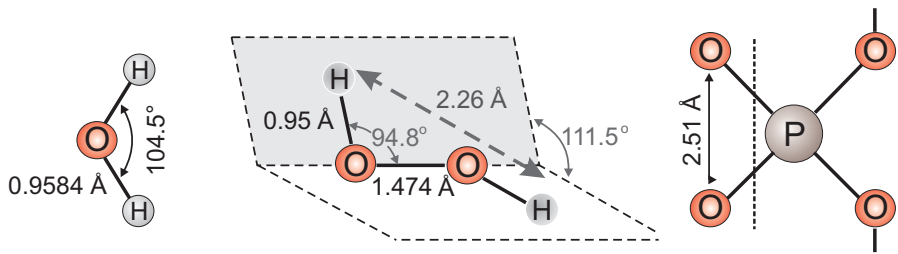}
}
\caption{Schematic structure of hydrogen peroxide molecule, water
molecule, and phosphate. Characteristic values of bond lengths,
valence angle and dihedral angles are showed.}
\label{fig:2}       
\end{figure}

The method of atom-atom potential functions is used for
calculation of energies of complexes ~\cite{Poltev1,Poltev2}.
Within the framework of this method the energy is presented as the
sum of pair interactions between atoms which belong to different
structure elements. The energy of complex can be presented as the
sum of two components:
\begin{equation}
U=\sum_{ij}\left[U_{el}(r_{ij})+U_{vdw}(r_{ij})\right],
\end{equation}
where $U_{el}(r_{ij})$ and $U_{vdw}(r_{ij})$ are the potential
energy of electrostatic and van der Waals interactions between $i$
and $j$ atoms at the distance $r_{ij}$.

Electrostatic interaction is defined as follows:
\begin{equation}
U_{el}(r_{ij})=\frac{q_iq_j}{4\pi\varepsilon\varepsilon_0r_{ij}},
\end{equation}
where $q_{i}$ and $q_{j}$ are the charges of $i$ and $j$ atoms,
$\varepsilon_{0}$ is the dielectric constant of vacuum,
$\varepsilon$ is the dielectric constant of medium. Charges of
atoms of hydrogen peroxide and water molecules are determined
using the values of dipole moments 2,1 D (H$_2$O$_2$) and 1,86 D
(H$_2$O)~\cite{Wikipedia}. Charges of atoms in phosphate group are
taken only for oxygen atoms in order to equate total charge with
-$e$ (Table 1). The value of dielectric constant $\varepsilon=1$
is used in calculations.

In the case of van der Waals interaction the Lennard-Jones
potential is used:
\begin{equation}
U_{vdw}(r_{ij})=-A_{ij}r_{ij}^{-6}+B_{ij}r_{ij}^{-12},
\end{equation}
where $A_{ij}$ and $B_{ij}$ are the constants (Table 1). The first
and the second terms in (3) describe the van der Waals repulsion
and short range repulsion between atoms, respectively. To take
into consideration the hydrogen bonds (O--H$\cdot\cdot\cdot$O) the
interaction between respective oxygen atoms are calculated using
the modified potential ~\cite{Poltev2}, where exponent in the
first term of equation (3) is changed to -10. Such exponent
describes the short-range character of hydrogen bonds.

For the description of interaction between ion and charged oxygen
atoms the Born-Mayer potential is used:
\begin{equation}
U_{BM}(r_{ij})=\frac{q_iq_j}{4\pi\varepsilon\varepsilon_0r_{ij}}\left[1-\frac{br_{ij}}{r_0^2}\exp\left(
\frac{r_{0}-r_{ij}}{b}\right)\right],
\end{equation}
where $b$ is the repulsion constant, $r_{0}$ is the equilibrium
length (Table 1). This potential was developed to describe the ion
crystal energy, and it takes into account electrostatic attraction
and short-range repulsion~\cite{Kittel}. The potential (4) has
been adapted in~\cite{PV1,PV2} for the description of counterion
interactions with the phosphate groups of DNA double helix
backbone.

Using the formulae (1) -- (4) with parameters in the Table 1 the
energies of different complexes  may be calculated.

\section{Complexes of hydrogen peroxide with DNA}
\label{sec:3}

Within the framework of proposed method the energies of
two-component (H$_{2}$O$_{2}$--PO$_{4}^{-}$,
H$_{2}$O--PO$_{4}^{-}$, Na$^{+}$--PO$_{4}$,
H$_{2}$O$_{2}$--Na$^{+}$, H$_{2}$O--Na$^{+}$) and three-component
(H$_{2}$O$_{2}$--Na$^{+}$--PO$_{4}^{-}$,
H$_{2}$O--Na$^{+}$--PO$_{4}^{-}$) complexes are calculated. The
most stable configurations for each complex are found.

\emph{Two-component complexes.} The most stable state for the
complex Na$^{+}$--PO$_{4}^{-}$ is obtained in case when sodium ion
is localized at the distance 2 \AA\ from the center of O--O
segment of the phosphate group. The calculated anergy  is the
lowest among two-component complexes (Fig. 3). This complex
corresponds to neutralization of DNA phosphate group by counterion
under the natural conditions.

The most stable complex of sodium ion with hydrogen peroxide is
obtained in the case  of ion localization at distance 2.66 \AA\ to
the center of O--O bond with equal distances to H atoms on the
opposite side of the molecule (Fig. 3). In the case of
Na$^{+}$--H$_2$O the most stable complex is obtained for the
Na$^{+}$--O distance 2.74 \AA. Under the natural conditions water
molecules in complex with Na$^{+}$ are the part of ion hydration
shell. The hydrogen peroxide molecule may substitute the water in
the ion   hydration shell.

The equilibrium configurations of H$_2$O$_2$--PO$_4^{-}$ complex
is obtained in the case of symmetrical localization of H atoms of
hydrogen peroxide molecule with respect to the O--O segment in the
phosphate group (Fig. 3). The complex has minimal energy for the
distance 2.85 \AA\ between centers of O--O segments in the
molecule and in the phosphate group. In the case of
H$_2$O--PO$_4^{-}$ complex the equilibrium configuration is
obtained for the distance 2.9 \AA\ of water oxygen to O--O segment
in the phosphate group (Fig. 3).

The hydrogen bonds, which can be formed between H (water or
hydrogen peroxide) and O (phosphate group), do not influence
significantly the energy values. In the case of the strongest
hydrogen bond (O--H--O atoms are localized on the line) the
calculated energies are lower than energies of
H$_{2}$O$_{2}$--Na$^{+}$--PO$_{4}$ and
H$_{2}$O--Na$^{+}$--PO$_{4}$ complexes. In configurations with two
hydrogen bonds between atoms of H$_{2}$O$_{2}$ (H$_{2}$O) molecule
the O--H--O segment is bent more than 30$^{\circ}$, that makes
such hydrogen bonds very weak.

\begin{figure}
\centering
\resizebox{0.7\textwidth}{!}{%
  \includegraphics{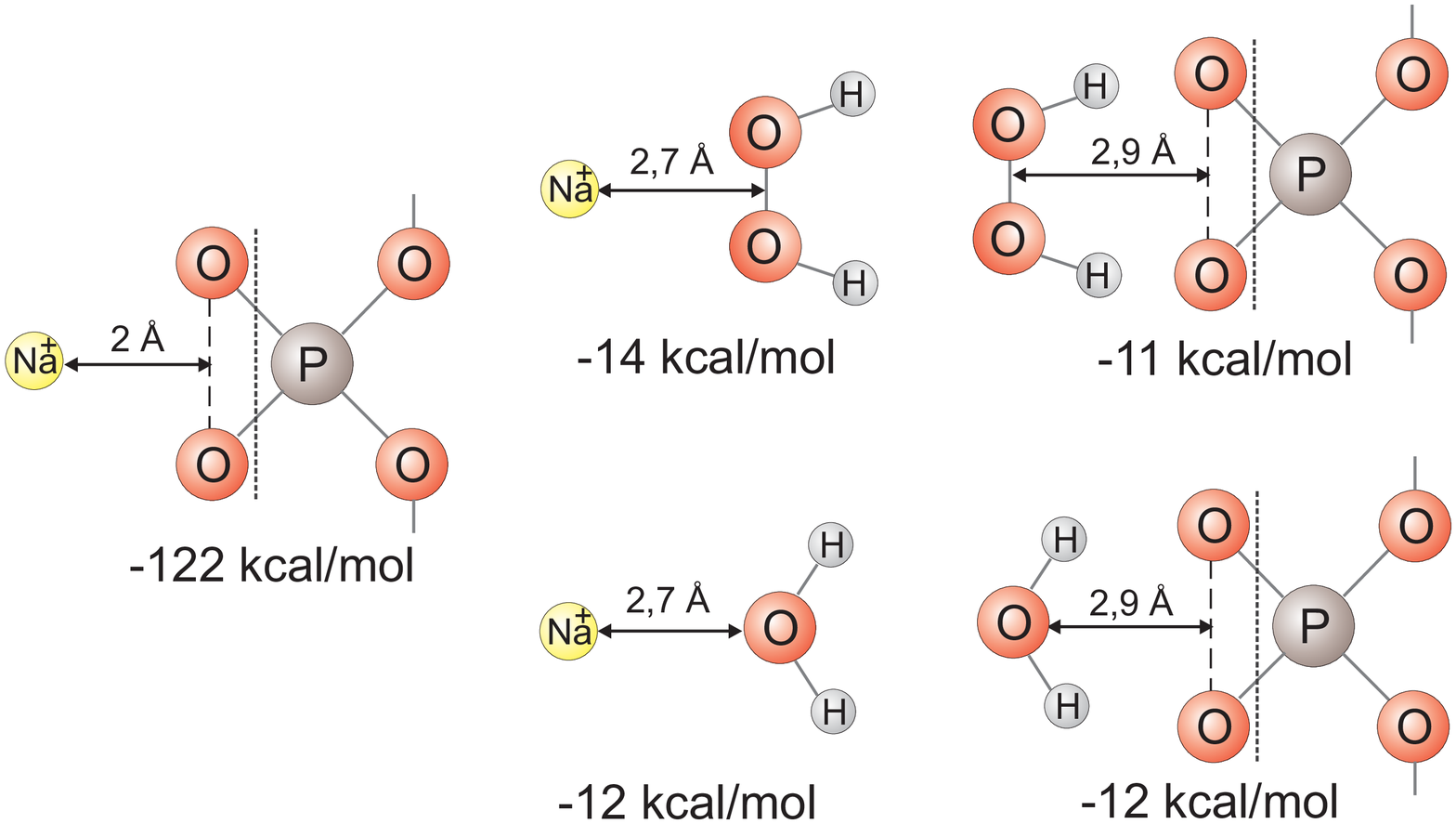}
}
\caption{The structures of two-component complexes:
Na$^{+}$--PO$_{4}$, H$_{2}$O$_{2}$--Na$^{+}$, H$_{2}$O--Na$^{+}$,
H$_{2}$O$_{2}$--PO$_{4}$, H$_{2}$O--PO$_{4}$ .}
\label{fig:3}       
\end{figure}

The equilibrium configurations of H$_2$O$_2$--PO$_4^{-}$ and
H$_2$O--PO$_4$ complexes depend  on mutual disposition of
molecules. Rotating H$_2$O$_{2}$ molecule with respect to the axis
passing through the center of O--O segment, the most stable
complex is found for the angle about $60^\circ$ (energy -12.8
kcal/mol) (Fig. 4a). In the case of molecule rotates around O--O
bond, the energy minimum is obtained for the angle $0^\circ$,
while at $180^\circ$ the repulsion occurs (Fig. 4b). Some small
minimums at about the angles $70^\circ$ and $290^\circ$ appear due
to the van der Waals repulsion and Coulomb attraction (dash lines
at figure). In the case of rotation of water molecule the obtained
results are qualitatively similar.

\begin{figure}
\centering
\resizebox{0.7\textwidth}{!}{%
  \includegraphics{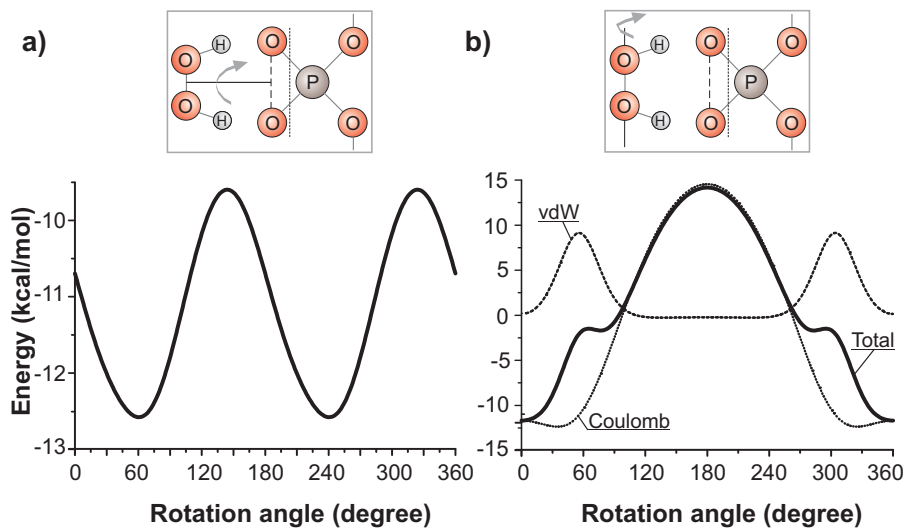}
}
\caption{Energy of  H$_2$O$_2$--PO$_4^{-}$ complex as a function
of rotation angle of H$_2$O$_2$ molecule. a) The dependence of
energy on rotation angle with respect to the axis passing through
the centers of  O--O segments in H$_2$O$_2$ and PO$_4$ (is showed
in inset). b) The dependence of energy on rotation angle with
respect to the O--O bond of H$_2$O$_2$ molecule (is showed in
inset). The van der Waals and Colomb contributions to the energy
are showed by dashed lines.}
\label{fig:5}       
\end{figure}

\emph{Three-component complexes.} The calculations are carried out
for  H$_2$O$_2$--Na$^{+}$--PO$_4^{-}$ and
H$_2$O--Na$^{+}$--PO$_4^{-}$ complexes. The energy of the
complexes are calculated for different distances between hydrogen
peroxide (water) molecule and ion-phosphate complex. The
ion-phosphate distance is taken equal to 2 \AA, correspondingly to
the result for two-component complex Na$^{+}$--PO$_4$ (Fig. 3).
Such ion localization is the most energetically favorable,
therefore other variants of ion localization are not considered in
the present work.

The configurations where hydrogen atoms of H$_2$O$_2$ (H$_2$O)
molecule are turned toward and outward phosphate group are
considered (Fig. 5a). In the first case there is no energy minimum
due to the  electrostatic repulsion between hydrogen atoms and
sodium ion. Increasing the distance to H$_2$O$_2$ (H$_2$O) the
energy of the complex monotonously decreases to the value of
Na$^{+}$--PO$_4$ complex (Fig. 5b). In the case when H atoms of
H$_2$O$_2$ (H$_2$O) molecule are turned toward the phosphate group
the energy minimum –130.3 kcal/mol (–129.2 kcal/mol) appear  at
the distance 2.75 \AA\ (2.82 \AA). The minimum is conditioned by
the electrostatic attraction between ion and oxygen atoms of
molecule. Rotation of hydrogen peroxide or water molecules around
axis which passes through center of O--O segment plays
insignificant role because the total energy depends on interaction
between ion and oxygen atoms mainly. Thus, the stability of
H$_2$O$_2$--Na$^{+}$--PO$_4$ and H$_2$O--Na$^{+}$--PO$_4$
complexes are almost equal.

\begin{figure}
\centering
\resizebox{0.6\textwidth}{!}{%
  \includegraphics{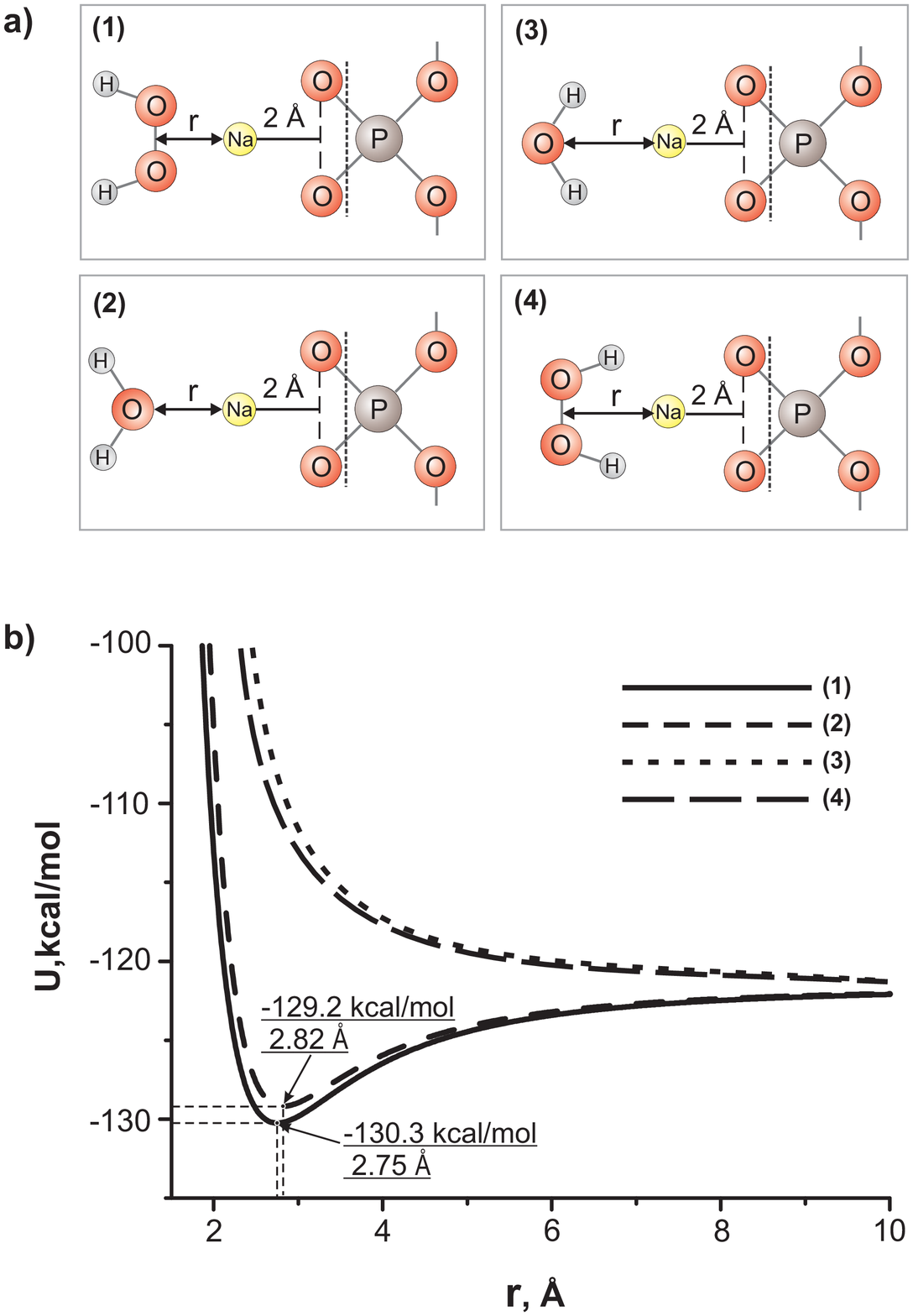} }
\caption{Energy of three-component complexes. a) Position of
H$_{2}$O$_{2}$ and  H$_{2}$O molecules in the considered
complexes. b) Energy of the complex  as a function of distance
between H$_{2}$O$_{2}$ or H$_{2}$O molecules to the ion-phosphate
complex.}
\label{fig:4}       
\end{figure}

\section{Vibrations of hydrogen peroxide in DNA low-frequency Raman spectra}

The formation of stable complexes of H$_{2}$O$_{2}$ molecule with
DNA phosphate groups should be accompanied by the appearance of
specific modes of vibrations  of hydrogen peroxide molecule in the
vibrational spectra. The frequencies of vibrations of
H$_{2}$O$_{2}$--Na$^{+}$--PO$_{4}^{-}$ complex are expected to be
in the low-frequency spectra range ($<$200 cm$^{-1}$),
analogically to the modes of counterion vibrations with respect to
DNA phosphate groups (ion-phosphate
modes)~\cite{PV1,PV2,PV3,BVKP}. The low-frequency spectra of DNA
is known to be characterized by the conformational vibrations of
the double helix structural elements as whole (nucleosides,
nucleotides, phosphate groups)~\cite{VK1,VK2,VK3,VK4}. The
vibrations of hydrogen peroxide molecule with respect to the DNA
phosphate groups should disturb the internal dynamics of the
double helix changing the low-frequency spectra shape.

To determine the influence of H$_{2}$O$_{2}$ molecules on DNA
conformational vibrations   the phenomenological model describing
the Raman low-frequency spectra of DNA with
counterions~\cite{PV1,PV2,PV3} is used. Within the framework of
this model the DNA double helix is represented as the double chain
of nucleotides with counterions tethered to the phosphate groups
of the macromolecule backbone. To take into consideration the
influence of hydrogen peroxide molecule this model is modified by
increasing the mass of tethered counterions for the mass of one
H$_{2}$O$_{2}$ molecule. The low-frequency vibration spectra of
DNA with counterions is also calculated for the case of H$_{2}$O
molecule attached to the ion. The equations, necessary for the
calculations of the mode frequencies and Raman intensities, are
the same as in the original works~\cite{PV2,PV3}. The calculation
parameters are taken for the case of \emph{B}-form of the double
helix.

\begin{figure}[b!]
\centering
\resizebox{0.5\textwidth}{!}{%
  \includegraphics{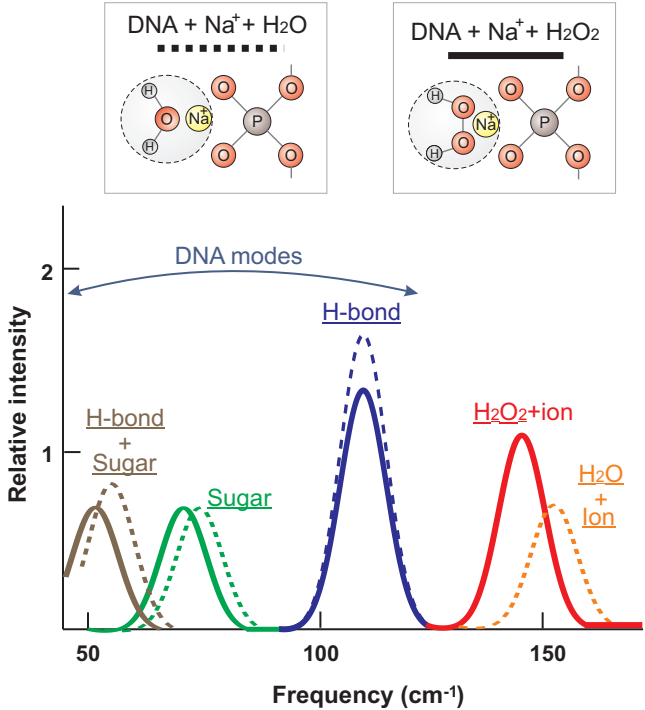}
}
\caption{The low-frequency Raman spectra of DNA with
H$_{2}$O$_{2}$--Na$^{+}$ complex (solid line) and  with
H$_{2}$O--Na$^{+}$ complex (dashed line). In the insets the
structure of respective complexes are showed.}
\label{fig:1}       
\end{figure}

As the result the low-frequency Raman spectra of DNA with attached
H$_{2}$O$_{2}$-Na$^{+}$ complex is calculated as well as the
spectra of DNA with H$_{2}$O-Na$^{+}$ complex (Fig. 6). The
internal vibrations of the double helix (lower than 120 cm$^{-1}$)
are connected with the vibrations of H-bond stretching in nucleic
bases and the vibrations in nucleosides due to the flexibility of
sugar ring. These modes do not sense essentially the presence of
hydrogen peroxide molecule. At the higher frequencies the modes,
characterizing the vibrations of the complexes with respect to the
phosphate groups, are determined about 155 cm$^{-1}$ and 140
cm$^{-1}$, respectively. The mode intensity is higher in the case
of complex with hydrogen peroxide molecule. Thus, our estimates
show that hydrogen peroxide molecule, attached to DNA phosphate
groups, may induce the visible changes in the low-frequency Raman
spectra.

\section{Discussion}

The calculations show that the complexes including hydrogen
peroxide and water molecules are the most stable under the
presence of sodium counterion. Without Na$^{+}$ the stability of
the complexes decreases for about one order. That is due to the
dominant role of electrostatic interactions in the complex
formation. The stability of H$_{2}$O$_{2}$--Na$^{+}$--PO$_{4}^{-}$
is about the same as H$_{2}$O--Na$^{+}$--PO$_{4}^{-}$. Therefore,
hydrogen peroxide molecule may be expected to incorporate into the
hydration shell of sodium counterion already tethered to the
oxygen atoms of DNA phosphate group.

The formation of stable complex of H$_{2}$O$_{2}$--Na$^{+}$ with
the phosphate group may be detected by  the low-frequency Raman
spectra of DNA. According to our estimations the vibrations  of
counterion with hydrogen peroxide molecule with respect to the
phosphate group of the double helix backbone should be about 10
cm$^{-1}$ lower than the frequencies of vibrations of the complex
with water molecule. Taking into consideration the determined
frequency shift the lifetimes of
H$_{2}$O$_{2}$--Na$^{+}$--PO$_{4}^{-}$ and
H$_{2}$O--Na$^{+}$--PO$_{4}^{-}$ complexes may be compared.
According to the Arrhenius equation the ratio between
characteristic lifetimes may be written in the following form:

\begin{equation}
\frac{\tau_{1}}{\tau_{2}}=\frac{\omega_{2}}{\omega_{1}}\exp\left(\frac{E_{1}-E_{2}}{kT}\right)\approx
\frac{\omega_{2}}{\omega_{1}},
\end{equation}
where $\tau_{1}$, $\omega_{1}$ and $\tau_{2}$, $\omega_{2}$ are
the lifetimes and frequencies of the complex with water and
hydrogen peroxide molecules, respectively; $E_{1}$ and $E_{2}$ are
the energies of the complexes with hydrogen peroxide and water
molecules, respectively. The energies of the considered complexes
are rather close (Fig. 5), therefore the value of the energy
difference can be equal to zero. As the result the expected
lifetime of the complex with H$_{2}$O$_{2}$ is higher than in the
complex with H$_{2}$O molecule.

Due to the larger lifetime of
H$_{2}$O$_{2}$--Na$^{+}$--PO$_{4}^{-}$ complex the hydrogen
peroxide molecules may be accumulated near DNA double helix. At
significant amounts the H$_{2}$O$_{2}$ molecules can influence DNA
biological functioning by two mechanisms. Firstly H$_{2}$O$_{2}$
can induce double strand breaks in double helix backbone by
decaying into radicals, and secondly H$_{2}$O$_{2}$ molecules may
break nucleic-protein recognition, stopping the processes of
translation of genetic information. Thus, the hydrogen peroxide
molecule, interacting with DNA double helix, can deactivate cell
genetic apparatus and induce cell death.

\section{Conclusions}
The hydrogen peroxide molecules appear in the cell medium after
the passage of high-energy ions. To study the possible effect of
H$_2$O$_2$  molecule in the functioning of the living cell the
complexes of hydrogen peroxide molecules with the phosphate groups
of DNA backbone has been considered. Using the method of atom-atom
potential functions the stability of different complexes,
consisting of DNA phosphate group, H$_2$O$_2$ (or H$_2$O)
molecule, and Na$^{+}$ metal ion, have been studied. The results
show that the complexes with H$_2$O$_2$ molecule and phosphate
group of DNA are no less stable than respective complexes with
H$_2$O molecule. The sodium counterion neutralizing the charge of
phosphate group   plays the key role in stabilization of these
complexes. For the case of most stable complexes of hydrogen
peroxide with the phosphate groups the low-frequency Raman spectra
of DNA have been calculated. The vibrations of the complex as
whole with respect to DNA phosphate group has been found near 150
cm$^{-1}$. The observation of this mode may be considered as a
test revealing the complex formation. Taking into consideration
the determined frequencies of vibrations the lifetimes of
H$_{2}$O$_{2}$--Na$^{+}$--PO$_{4}^{-}$ and
H$_{2}$O--Na$^{+}$--PO$_{4}^{-}$ complexes have been compared, and
the lifetime of  the complex with hydrogen peroxide has been found
higher  than in the case of water molecule. Taking this into
consideration the hydrogen peroxide molecules are expected to be
accumulated near DNA macromolecule. The action of H$_2 $O$_2$ on
DNA structure and dynamics may be sufficient for the braking the
processes of DNA biological functioning. The influence of hydrogen
peroxide molecules on DNA can be regarded as an individual channel
in heavy ion therapy.

\end{document}